\documentstyle[12pt,psfig]{article}
\newcommand{\beq}{\begin{equation}}
\newcommand{\eeq}{\end{equation}}
\newcommand{\beqa}{\begin{eqnarray}}
\newcommand{\eeqa}{\end{eqnarray}}
\newcommand{\ba}{\begin{array}}
\newcommand{\ea}{\end{array}}

\newcommand{\ds}{\displaystyle}
\renewcommand{\thefootnote}{\fnsymbol{footnote}}

\makeatletter
\@addtoreset{equation}{section}

\makeatother

\begin{document}

\begin{titlepage}
\null
\begin{flushright}
UT-857
\\
hep-th/9909032
\\
September, 1999
\end{flushright}

\vskip 2cm
\begin{center}

  {\LARGE Geometry and $N=2$ Exceptional Gauge Theories}

\lineskip .75em
\vskip 2.5cm
\normalsize

  {\large Jiro Hashiba and Seiji Terashima}

\vskip 1.5em

  {\large \it Department of Physics, University of Tokyo\\
               Tokyo 113-0033, Japan}

\vskip 2cm

{\bf Abstract}
\end{center}

We find the Seiberg-Witten geometry for four dimensional $N=2$
 supersymmetric $E_6$ gauge theories with massless fundamental
 hypermultiplets, by geometrically embedding them in type II string
 theories compactified on Calabi-Yau threefolds. The resulting geometry
 completely agrees with that of recent works, which are based on the
 technique of $N=1$ confining phase superpotentials. We also derive the
 Seiberg-Witten geometry for $E_7$ gauge theories with massive
 fundamental hypermultiplets. 

\end{titlepage}

\renewcommand{\thefootnote}{\arabic{footnote}}
\baselineskip=0.7cm

\clearpage
%\tableofcontents
%\clearpage

%%%%%%%%%%%%%%%%%%%%%%%%%%%%%%%%%%%%%%%%%%%%%%%%%%%%%%%%%%%%%%%%%%%
%%%%%%%%%%%%%%%%%%%%%%%%%%%%%%%%%%%%%%%%%%%%%%%%%%%%%%%%%%%%%%%%%%%
\section{Introduction}
%%%%%%%%%%%%%%%%%%%%%%%%%%%%%%%%%%%%%%%%%%%%%%%%%%%%%%%%%%%%%%%%%%%
%%%%%%%%%%%%%%%%%%%%%%%%%%%%%%%%%%%%%%%%%%%%%%%%%%%%%%%%%%%%%%%%%%%

In the past few years, there has been much development in our understanding
of non-perturbative properties of supersymmetric gauge theories and
superstring theories. On one hand, it has been found that exact results
for the Coulomb branch of $N=2$ gauge theories in four dimensions can be 
obtained by considering auxiliary Riemann surfaces \cite{SW}. On the
other hand, it has been also recognized that D-branes play the role of
solitonic objects, and realize enhanced gauge symmetries especially in
type II string theories compactified on singular manifolds
\cite{KMP}. It would be tempting to put these together, namely to embed
$N=2$ gauge theories in string theories and conjecture that the Riemann
surfaces on field theory side originate from compactifying manifolds on
string theory side.

Calabi-Yau threefold compactification of type II string theories indeed
provides a systematic way of finding exact solutions of four dimensional 
$N=2$ supersymmetric gauge theories \cite{KKV,KMV,Mayr}. However, a low
energy effective theory of type II string theories contains not only
gauge theory degrees of freedom, but also a gravity multiplet. In
Calabi-Yau compactification approach, the gauge theory fields propagate
near the singularities of the Calabi-Yau, as contrasted to the gravity
multiplet which propagates on the entire Calabi-Yau space. We can
therefore decouple gravity effects and consider pure gauge theories by
focusing on the vicinity of the Calabi-Yau singularities.

If we construct a gauge theory from type IIA string theory, the Coulomb
branch of the gauge theory is identified with the K\"{a}hler moduli of
the compactifying Calabi-Yau. In type IIA theory, the K\"{a}hler moduli
receive quantum corrections due to worldsheet instanton effects. So,
one-loop and non-perturbative corrections to the Coulomb branch are not
computable from type IIA perspective. This defect can be remedied by
mapping the K\"{a}hler moduli of the type IIA Calabi-Yau to the complex
moduli of the {\it mirror} Calabi-Yau compactifying type IIB
theory. Since the complex moduli of the type IIB Calabi-Yau is free from
quantum corrections, the exact metric on the Coulomb branch can be
obtained from {\it classical} type IIB Calabi-Yau geometry.

In this article, we concentrate on $E_6$ and $E_7$ gauge theories with
fundamental matter. For the $E_6$ case, the Seiberg-Witten geometry with
massless fundamental hypermultiplets is derived from the mirror symmetry
between type IIA and type IIB string theories. The resulting geometry
coincides with that obtained in \cite{TY}. For the $E_7$ case, we apply
various decoupling limits to the gauge theory, to obtain the geometry
with massive fundamental hypermultiplets. The result is again
consistent with \cite{Brodie}, which have presented the geometry with a
massless half hypermultiplet. In both cases, the Seiberg-Witten 
geometry has the form of an ALE fibration over 2-sphere, with the
fibration data being slightly complicated due to the existence of extra
matter. The appearance of ALE fibration is not specific to the present
$E_6$ and $E_7$ cases. In general, the vector moduli information of an
$N=2$ gauge theory is expected to be more naturally encoded in an ALE
fibration over 2-sphere, than in a one complex dimensional space
\cite{KLMVW,LW}. $N=2$ field theories with other exceptional and some
$SO(N)$ gauge groups are discussed in \cite{Brodie,AG}.

This paper is organized as follows. In section 2, we derive the
Seiberg-Witten geometry for $E_6$ gauge theories with massless
fundamental matter, by geometrically realizing them on Calabi-Yau
singularities. In section 3, the Seiberg-Witten geometry is presented
for $E_7$ gauge theories with massive fundamental matter. As a
by-product, we will also find the geometry for $SO(12)$ gauge theories
with massive fundamentals and spinors. The last section is devoted to
discussion and conclusions.

%%%%%%%%%%%%%%%%%%%%%%%%%%%%%%%%%%%%%%%%%%%%%%%%%%%%%%%%%%%%%%%%%%%
%%%%%%%%%%%%%%%%%%%%%%%%%%%%%%%%%%%%%%%%%%%%%%%%%%%%%%%%%%%%%%%%%%%
\section{Geometric construction of $E_6$ gauge theories}
%%%%%%%%%%%%%%%%%%%%%%%%%%%%%%%%%%%%%%%%%%%%%%%%%%%%%%%%%%%%%%%%%%%
%%%%%%%%%%%%%%%%%%%%%%%%%%%%%%%%%%%%%%%%%%%%%%%%%%%%%%%%%%%%%%%%%%%

In this section, we will investigate a mirror pair of type
IIA and type IIB string theories compactified on Calabi-Yau threefolds. 
Type IIA string theory has an advantage that gauge groups and matter
representations are easily identified. However, the Seiberg-Witten
geometry can be directly obtained from the Calabi-Yau compactifying type
IIB string theory. These are why we deal with both type IIA and type
IIB string theories, not only with one of them. Throughout this section,
toric geometry will play a crucial role. For details of toric geometry
and its application to mirror symmetry, the reader should consult
references \cite{HKTY,Batyrev,AGR,Greene}, where sections 9 and 10
of Greene's review \cite{Greene} contain a basic introduction.

We consider first type IIA string theory compactified on a Calabi-Yau
threefold $X$, which is a $K3$ fibration over ${\bf P^1}$
surface. Furthermore, we assume that the $K3$ fiber itself is an
elliptic fibration over ${\bf P^1}$. Therefore the Calabi-Yau $X$ can
also be regarded as an elliptic fibration over the Hirzebruch surface
${\bf F_n}$, where the integer $n$ determines how ${\bf P^1}$ is fibered
over ${\bf P^1}$ in ${\bf F_n}$. Type IIA string theory on $X$ in this
particular class is conjectured to be dual to $E_8 \times E_8$ heterotic
string theory on $K3 \times T^2$, with no Wilson line turned on $T^2$
\cite{KachVa,Aspinwall}. It is then possible to consistently take a
large volume limit of $T^2$, and lift the duality to the six dimensional
one. This heterotic dual description in six dimensions strongly suggests
that the gauge symmetries and matter representations we will determine
in the following are correct \cite{BIKMSV}.

In order for the gauge theory resulting from this compactification of
type IIA string theory to possess $E_6$ gauge symmetry, the $K3$ fiber
has to develop an $E_6$ type singularity. This requirement is equivalent
to that the elliptic fiber degenerates as one approaches some point on
the ${\bf P^1}$ base of the $K3$ fiber, with the degeneration being of
$E_6$ type. Such a singular Calabi-Yau $X$ can be embedded in a ${\bf
WP_{1,2,3}^2}$ bundle over ${\bf F_n}$. Let $(x,y,z),~(s,t)$, and
$(u,v)$ be the homogeneous coordinates on ${\bf WP_{1,2,3}^2}$, the
${\bf P^1}$ fiber of ${\bf F_n}$, and the ${\bf P^1}$ base of ${\bf
F_n}$, respectively. The weights of them are as follows:
\begin{equation}
  \label{weight}
  \begin{array}{ccccccc}
       x &    y & z & s & t & u & v, \\
       2 &    3 & 1 & 0 & 0 & 0 & 0, \\
       4 &    6 & 0 & 1 & 1 & 0 & 0, \\
    2n+4 & 3n+6 & 0 & n & 0 & 1 & 1.
  \end{array}
\end{equation}
Then, $X$ is written as the hypersurface equation \cite{BIKMSV}
\begin{equation}
  \label{weierstrass}
    y^2 = x^3 + f(s,t;u,v)xz^4 + g(s,t;u,v)z^6,
\end{equation}
where 
\begin{equation}
  \label{polynomials}
    f(s,t;u,v) = \sum_{i=3}^I s^i t^{8-i}f_{8+n(4-i)}(u,v),~~~~
    g(s,t;u,v) = \sum_{j=4}^J s^j t^{12-j}g_{12+n(6-j)}(u,v).
\end{equation}
In (\ref{polynomials}), $f_{8+n(4-i)}$ and $g_{12+n(6-j)}$ are
homogeneous polynomials of degrees specified by their
subscripts. The indices $I$ and $J$ denote the largest values of $i$ and
$j$ such that all the degrees appearing in (\ref{polynomials}) are not
negative. The fact that $f(s,t;u,v)$ and $g(s,t;u,v)$ are divisible by
$s^3$ and $s^4$ guarantees that the $K3$ fiber has an $E_6$ singularity,
provided that the polynomials $f_{8+n(4-i)}$ and $g_{12+n(6-j)}$ have
generic coefficients.

There are a few comments on the type IIA Calabi-Yau $X$ given by
(\ref{weierstrass}) with (\ref{polynomials}). First, there may exist
some singularities other than the $E_6$ singularity in which we are
now interested, when the polynomials $f_{8+n(4-i)}$ and
$g_{12+n(6-j)}$ have some special forms. If this occurs, some extra gauge
theories will arise from the other singularities. We have then to decouple
somehow the extra gauge theories from our $E_6$ gauge theory. We will
argue this subtlety later in this section. Second, in addition to the
$E_6$ vector multiplet, the Calabi-Yau $X$ automatically incorporates
$N_f \equiv n+6$ hypermultiplets in fundamental representation
\cite{BIKMSV}. The hypermultiplets are localized at $N_f$ extra
singularities on the ${\bf P^1}$ base of ${\bf F_n}$, i.e., the points
where the $E_6$ singularity of the $K3$ fiber becomes worse
\cite{BIKMSV,KV,Tani}. Hereafter, we will assume $-6 \leq n < -2$ to
ensure that the gauge theory is asymptotically free, and $N_f \geq
0$. The lower bound $n \geq -6$ is also required from the restriction
that the $E_6$ gauge group must not be enhanced to larger gauge
groups. The upper bound $n < -2$ has another geometrical meaning, as we
will see later.

One can resolve the $E_6$ singularity of the $K3$ fiber by blowing 
up the ambient space, the ${\bf WP_{1,2,3}^2}$ bundle, at the locus
$\{x=y=s=0\}$. The smooth Calabi-Yau $X$ obtained by the resolution
admits a toric description, which is given by a polyhedron $\Delta$
whose vertices consist of the following vectors \cite{CF,BIKMSV},
\begin{equation}
  \label{vertices}
  \begin{array}{rcl}
     v_0 &=& (0,0,0,0), \\
     v_1 &=& (1,0,0,0), \\
     v_2 &=& (0,1,0,0), \\
     v_3 &=& (-2,-3,0,0), \\
     v_4 &=& (0,0,1,0), \\
     v_5 &=& (-4,-6,-1,0), \\
     v_6 &=& (0,0,0,1), \\
     v_7 &=& (-2n-4,-3n-6,-n,-1), \\
     v_8 &=& (2,3,1,0), \\
     v_9 &=& (3,5,2,0), \\
     v_{10} &=& (4,6,3,0), \\
     v_{11} &=& (3,4,2,0), \\
     v_{12} &=& (2,2,1,0), \\
     v_{13} &=& (2,3,2,0).
  \end{array}
\end{equation}
As is well known in toric geometry, each vertex $v_i$ in (\ref{vertices}) is
associated with a divisor $D_i$ in $X$. The three vertices $v_1,v_2$,
and $v_3$ describe the divisors in the ${\bf WP_{1,2,3}^2}$ fiber,
$\{x=0\},\{y=0\}$, and $\{z=0\}$ restricted on $X$. Similarly,
$v_4,v_5,v_6$, and $v_7$ are identified with the divisors in the ${\bf
F_n}$ base in $X$, $\{s=0\},\{t=0\},\{u=0\}$, and $\{v=0\}$. The six
vertices $v_8,\cdots,v_{13}$ represent the six exceptional divisors
arising from the resolution (as we will see below, their restriction on
the $K3$ fiber constitute six ${\bf P^1}$ surfaces, whose intersection
form is nothing but the $E_6$ Cartan matrix). Finally, notice that the
polyhedron $\Delta$ contains only one vector $v_0$ as its interior
point. In general, for the hypersurface defined by a polyhedron to be
Calabi-Yau, it is necessary that the polyhedron has a unique interior
point \cite{Batyrev}. The unique interior point corresponds to the
canonical divisor of the ambient toric variety.

Our next task is to present type IIB Calabi-Yau $\tilde{X}$, which is
the mirror partner of $X$. The Calabi-Yau $\tilde{X}$ takes the form of
hypersurface equation
\begin{equation}
  \label{mirror}
    \sum_{i=0}^{13} a_i y_i = 0,
\end{equation}
where the complex numbers $a_i$ parametrize the complex deformation
moduli of $\tilde{X}$. The complex variables $y_i$ are not independent
of each other but obey the constraints
\begin{equation}
  \label{constraint}
    \prod_{i=0}^{13} y_i^{l_i^{(a)}} = 1,
\end{equation}
where $l^{(a)}$ are fourteen dimensional vectors such that the
following linear relations hold:
\begin{equation}
  \label{relation}
    \sum_{i=0}^{13} l_i^{(a)} (v_i,1) = 0.
\end{equation}
Here, $(v_i,1)$ denotes the five dimensional vector made by adding the
fifth component $1$ to the vector $v_i$. In the present case, the number
of independent $l^{(a)}$'s is nine, hence the index $a$ runs from 1 to 9
in (\ref{constraint}) and (\ref{relation}). The explicit form of
$l^{(a)}$ is given by
%\begin{equation}
%  \label{charge-vectors}
%    {\tiny
%    \begin{array}{rccrrrrrrrrrrrrrr}
%              & & (&v_0,&v_1,&v_2,&v_3,&v_4,&v_5,&v_6,&v_7;
%                   &v_8,&v_9,&v_{10},&v_{11},&v_{12},&v_{13}), \\
%      l^{(1)} &=& (&  -6,&2,&3,&1,&0,&0,&0,&0;&0, &0, &0, &0, &0, & 0), \\
%      l^{(2)} &=& (& -12,&4,&6,&0,&1,&1,&0,&0;&0, &0, &0, &0, &0, & 0), \\
%      l^{(3)} &=& (& -6n-12,&2n+4,&3n+6,&0,&n,&0,&1,&1;
%                    &0,&0,&0,&0,&0,&0), \\
%      l^{(4)} &=& (&  -1,&1,&1,&0,&0,&0,&0,&0;&-2,&1, &0, &0, &0, & 0), \\
%      l^{(5)} &=& (&  -1,&0,&1,&0,&0,&0,&0,&0;&1, &-2,&1, &0, &0, & 0), \\
%      l^{(6)} &=& (&  -1,&0,&0,&0,&0,&0,&0,&0;&0, &1, &-2,&1, &0, & 1), \\
%      l^{(7)} &=& (&   0,&0,&0,&0,&0,&0,&0,&0;&0, &0, &1, &-2,&1, & 0), \\
%      l^{(8)} &=& (&   0,&1,&0,&0,&0,&0,&0,&0;&0, &0, &0, &1, &-2,& 0), \\
%      l^{(9)} &=& (&   0,&0,&0,&0,&1,&0,&0,&0;&0, &0, &1, &0, &0, &-2).
%    \end{array}
%    }
%\end{equation}
\begin{equation}
  \label{charge-vectors}
  \left(
    \begin{array}{ccc}
      l_0^{(1)}    & \cdots & l_0^{(9)}    \\
      \vdots       & \ddots & \vdots       \\
      l_{13}^{(1)} & \cdots & l_{13}^{(9)}
    \end{array}
  \right) =
  \left(
    \begin{array}{rrrrrrrrr}
      -6,&-12,&-6n-12,&-1,&-1,&-1,& 0,& 0,& 0 \\
       2,&  4,&  2n+4,& 1,& 0,& 0,& 0,& 1,& 0 \\
       3,&  6,&  3n+6,& 1,& 1,& 0,& 0,& 0,& 0 \\
       1,&  0,&     0,& 0,& 0,& 0,& 0,& 0,& 0 \\
       0,&  1,&     n,& 0,& 0,& 0,& 0,& 0,& 1 \\
       0,&  1,&     0,& 0,& 0,& 0,& 0,& 0,& 0 \\
       0,&  0,&     1,& 0,& 0,& 0,& 0,& 0,& 0 \\
       0,&  0,&     1,& 0,& 0,& 0,& 0,& 0,& 0 \\
       0,&  0,&     0,&-2,& 1,& 0,& 0,& 0,& 0 \\
       0,&  0,&     0,& 1,&-2,& 1,& 0,& 0,& 0 \\
       0,&  0,&     0,& 0,& 1,&-2,& 1,& 0,& 1 \\
       0,&  0,&     0,& 0,& 0,& 1,&-2,& 1,& 0 \\
       0,&  0,&     0,& 0,& 0,& 0,& 1,&-2,& 0\\
       0,&  0,&     0,& 0,& 0,& 1,& 0,& 0,&-2 \\
    \end{array}
  \right)
\end{equation}
Note that the weights of the ${\bf WP_{1,2,3}^2}$ bundle (\ref{weight})
appears in the $3 \times 7$ entries
$l_i^{(a)}~(a=1,2,3;i=1,\cdots,7)$. Moreover, in (\ref{charge-vectors})
the $E_6$ Cartan matrix has emerged as the $6 \times 6$ components
$l_i^{(a)}~(a=4,\cdots,9;i=8,\cdots,13)$. In general, each linear
relation $l^{(a)}$ corresponds to a curve class $C^a$ in $X$. It is also
well known that the component $l_i^{(a)}$ is proportional to the
intersection number $C^a \cdot D_i$. In the present analysis,
$l^{(4)},\cdots,l^{(9)}$ are identified with the six blown up 2-spheres
in the $K3$ fiber,
\footnote{\label{identification}
$l^{(1)},l^{(2)}$, and $l^{(3)}$ are identified with three 2-cycles in
$X$, namely the elliptic fiber, the ${\bf P^1}$ fiber of
${\bf F_n}$, and the ${\bf P^1}$ base of ${\bf F_n}$, respectively.
Here, the ``${\bf P^1}$ base of ${\bf F_n}$'' actually means the
singular locus in $X$, $\{ x=y=s=0 \}$.
} and $D_8,\cdots,D_{13}$ correspond to the six exceptional divisors in
$X$. If we restrict these exceptional divisors on the $K3$ fiber
of $X$, they become identical to the six blown up 2-spheres. Therefore,
the appearance of the $E_6$ Cartan matrix ensures that the six
exceptional ${\bf P^1}$'s in the $K3$ fiber lead to $E_6$ gauge
symmetry. 
%In Figure 2, we show the vertices for type IIA non-compact $K3$ fiber.  
%\begin{equation}
%  v_i^T \rightarrow \left(
%                    \begin{array}{cccc}
%                      0 & 0 & 0 & 0 \\
%                      0 & 0 & 0 & 0 \\
%                      0 & 0 & 0 & 0 \\
%                      0 & 0 & 0 & 1
%                    \end{array} \right) v_i^T + \left(
%                    \begin{array}{c}
%                      1 \\
%                      1 \\
%                      1 \\
%                      0
%                    \end{array} \right)
%\end{equation}
%\begin{figure}
%   \centerline{\psfig{figure=e6toric.eps,width=11cm}}
%   \caption{Toric vertices for type IIA non-compact $K3$ fiber.}
%\end{figure}

Let us now turn to examining the type IIB hypersurface (\ref{mirror}).
Because of the relation $\sum_{i=0}^{13}l_i^{(a)}=0$ following from
(\ref{relation}), both of the equations (\ref{mirror}) and
(\ref{constraint}) are invariant under the rescaling $y_i \rightarrow
\lambda y_i,~\lambda \in {\bf C}^*$. It is thus allowed to scale $y_i$'s
and put one of them to unity. We take here $y_{10}=1$. Then, the
constraints (\ref{constraint}) can be solved by the four independent
variables $x_1 \equiv y_9,x_2 \equiv y_{11},x_3 \equiv y_{13}$, and
$\zeta \equiv y_6$, that is, all other $y_i$'s can be represented by
some powers of $x_1,x_2,x_3$, and $\zeta$. The type IIB mirror manifold
(\ref{mirror}) is thus rewritten in terms of $x_1,x_2,x_3$, and $\zeta$
as
\begin{equation}
  \label{mirror2}
  \begin{array}{rcl}
    W &\equiv& \zeta + a_7 {\ds \frac{{x_3}^{N_f}}{\zeta}} \\
            && + {x_2}^3 + {x_3}^2 + 2{x_1}^2x_3 + a_{12} {x_2}^2 + a_8 x_1x_2
               + a_{13} x_3 + a_{11} x_2 + a_9 x_1 + a_{10} \\
            && + {\ds \frac{1}{{M_5}^{12}}}{x_3}^4
               + {\ds \frac{1}{{M_3}^6}}{x_3}^3
               + {\ds \frac{1}{M_0}}x_1x_2x_3 = 0,
  \end{array}
\end{equation}
where we have put $a_1=a_4=a_6=1$ and $a_2=2$, taking into account the
rescaling degrees of freedom of $x_1,x_2,x_3$, and $\zeta$. We have also
defined scale parameters $M_0,M_3$, and $M_5$ by $a_0 \equiv
\frac{1}{M_0},a_3 \equiv \frac{1}{{M_3}^6}$, and $a_5 \equiv
\frac{1}{{M_5}^{12}}$. In order for the second line in the r.h.s. of
(\ref{mirror2}) to become the standard form of versal deformation of
the $E_6$ singularity, one must suitably reparametrize $x_1,x_2$, and
$x_3$, and take the limit $M_0,M_3,M_5 \rightarrow \infty$. Let us
introduce new variables
\renewcommand\arraystretch{1.7}
\begin{equation}
  \label{reparam1}
  \begin{array}{rcl}
     x &\equiv& -(x_2 + c), \\
     y &\equiv& -i\left( x_3 + x_1^2 + {\ds \frac{1}{2}}a_{13}
                + {\ds \frac{1}{2M_0}}x_1x_2 \right), \\
     z &\equiv& x_1 + b,
  \end{array}
\end{equation}
where $b$ and $c$ are some constants determined later. Substituting
certain combinations of $x,y$, and $z$ into $x_1,x_2$, and $x_3$ by
means of (\ref{reparam1}), we obtain
\begin{equation}
  \label{mirror3}
  \begin{array}{rcl}
    W &=& \zeta + a_7 {\ds \frac{{x_3}^{N_f}}{\zeta}} \\
      & & -(y^2 + x^3 + z^4 + w_2 xz^2 + w_5 xz + w_6 z^2 
          + w_8 x + w_9 z + w_{12}) \\
      & & - {\ds \frac{1}{4{M_0}^2}} x^2z^2 + {\ds \frac{1}{M_0}} xz^3
          + {\ds \frac{1}{2{M_0}^2}}b x^2z \\
      & & + \left(4b + {\ds \frac{1}{M_0}}c\right)z^3
          + \left(a_{12} - {\ds \frac{1}{4{M_0}^2}}b^2 -3c\right)x^2 \\
      & & + {\ds \frac{1}{{M_5}^{12}}}{x_3}^4
          + {\ds \frac{1}{{M_3}^{6}}}{x_3}^3,
  \end{array}
\end{equation}
where it must be kept in mind that $x_3$ should be replaced with
\begin{equation}
  \label{x3}
    x_3 = iy - z^2 - \frac{1}{2M_0}bx + \left( 2b + \frac{1}{2M_0}c \right)z
          - \frac{1}{2}a_{13} - b^2 - \frac{1}{2M_0}bc +\frac{1}{2M_0}xz.
\end{equation}
The deformation parameters $w_i~(i=2,5,6,8,9,12)$ in (\ref{mirror3}) are
given by
\begin{equation}
  \label{deformation}
  \begin{array}{rcl}
    w_2 &=& {\ds \frac{3}{M_0}}b + {\ds \frac{1}{2{M_0}^2}}c, \\
    w_5 &=& a_8 - {\ds \frac{1}{2M_0}}a_{13}
            - {\ds \frac{3}{M_0}}b^2 - {\ds \frac{1}{{M_0}^2}}bc, \\
    w_6 &=& a_{13} + 6b^2
            + {\ds \frac{3}{M_0}}bc + {\ds \frac{1}{4{M_0}^2}}c^2, \\
    w_8 &=& a_{11} - a_8b - 2a_{12}c + 3c^2+ {\ds \frac{1}{2M_0}}a_{13}b
            + {\ds \frac{1}{M_0}}b^3 + {\ds \frac{1}{2{M_0}^2}}b^2c, \\
    w_9 &=& -a_9 - 2a_{13}b + a_8c - 4b^3 - {\ds \frac{1}{2M_0}}a_{13}c
            - {\ds \frac{3}{M_0}}b^2c - {\ds \frac{1}{2{M_0}^2}}bc^2, \\
    w_{12} &=& -a_{10} + {\ds \frac{1}{4}}{a_{13}}^2 + a_9b + a_{11}c
               + a_{13}b^2 - a_8 bc - a_{12} c^2 + c^3 + b^4 \\
           & & + {\ds \frac{1}{2M_0}}a_{13}bc + {\ds \frac{1}{M_0}}b^3c
               + {\ds \frac{1}{4{M_0}^2}}b^2c^2.
  \end{array}
\end{equation}
We require $b,c,a_8,\cdots,a_{13}$ to depend on $M_0$ so that
$w_i~(i=2,5,6,8,9,12)$ are fixed at some constants in the limit $M_0
\rightarrow \infty$. We can make all terms in the third and fourth line of
(\ref{mirror3}) vanish in the limit $M_0 \rightarrow \infty$, by
demanding that $b,c$, and $a_{12}$ depend on $M_0$ in the following way:
\begin{equation}
  \label{M-dependence}
  \begin{array}{rcl}
    b &=& M_0 d_2 + d_3
          + {\cal O}\left({\ds \frac{1}{M_0}}\right), \\
    c &=& -4{M_0}^2d_2 - 4M_0d_3 + d_4
          + {\cal O}\left({\ds \frac{1}{M_0}}\right), \\
    a_{12} &=& -12{M_0}^2d_2 - 12M_0d_3 + 3d_4 + {\ds \frac{1}{4}}{d_2}^2
               + {\cal O}\left({\ds \frac{1}{M_0}}\right),
  \end{array}
\end{equation}
where $d_2,d_3$, and $d_4$ are arbitrary constants independent of $M_0$. Note
that $w_2 = d_2 + {\cal O}(1/M_0)$, as one can see from the first equation
in (\ref{deformation}). Surprisingly, all the coefficients entering in the
r.h.s. of (\ref{x3}) are kept finite in the limit $M_0 \rightarrow
\infty$, as far as (\ref{deformation}) and (\ref{M-dependence}) are
satisfied. Hence, while the second term in the r.h.s. of (\ref{mirror3})
gives rise to a non-vanishing contribution, the two terms proportional to
$1/{M_3}^6$ and $1/{M_5}^{12}$ disappear in the limit $M_3,M_5
\rightarrow \infty$.

Eventually, we end up with the Seiberg-Witten geometry for the $E_6$
gauge theory of the form 
\begin{equation}
  \label{swgeometry}
    \zeta + \frac{1}{\zeta}(\Lambda_{E_6})^{24-6N_f}
    [X_{E_6}^{{\bf 27}}(x,y,z;w)]^{N_f} - W_{E_6}(x,y,z;w) = 0,
\end{equation}
where we have introduced $\Lambda_{E_6}$, the dynamical scale for the
theory, by making the identification $a_7=(\Lambda_{E_6})^{24-6N_f}$.
The polynomials $W_{E_6}$ and $X_{E_6}^{{\bf 27}}$ are given by
\begin{equation}
  \label{e6_deformation}
    W_{E_6}(x,y,z;w) = y^2 + x^3 + z^4+ w_2 xz^2 + w_5 xz + w_6 z^2
                       + w_8 x + w_9 z + w_{12},
\end{equation}
and
\begin{equation}
  \label{fundamental}
    X_{E_6}^{{\bf 27}}(x,y,z;w) = iy - z^2 - \frac{1}{2}w_2x - \frac{1}{2}w_6.
\end{equation}
The expression (\ref{fundamental}), which controls the
ALE fibration data for the theories with fundamental matter, was
derived from (\ref{x3}), (\ref{deformation}), and
(\ref{M-dependence}). The resultant manifold (\ref{swgeometry}) with
(\ref{e6_deformation}) and (\ref{fundamental}) completely agrees with
that obtained in \cite{TY}, which has been derived from the technique of 
$N=1$ confining phase superpotentials \cite{ElFoGiInRa, KiTeYa, TeYa1}.

Before closing this section, it is instructive to explain here why we had
to take the limit $M_0,M_3,M_5 \rightarrow \infty$, to obtain the
correct answer. As pointed out above, we must pay attention to the
vicinity of the $E_6$ singularity of the type IIA $K3$ fiber, in order
to appropriately ignore the effects from some other gauge
theories. Since the $E_6$ singularity is located at the locus
$\{x=y=s=0\}$ in $X$, the divisors $\{z=0\}$ and $\{t=0\}$ which
correspond to the vertices $v_3$ and $v_5$ in (\ref{vertices}) are
distant from the singularity. On the contrary, all other divisors in the
$K3$ fiber (which are given by the vertices
$v_1,v_2,v_4,v_8,\cdots,v_{13}$) intersect at least one of the
exceptional divisors appearing from the singularity. Thus, we may
eliminate the vertices $v_3$ and $v_5$ from (\ref{vertices}), to
``forget about'' other gauge theories, without affecting the $E_6$ gauge 
theory localized at the $E_6$ singularity. In the present mirror map,
the divisors $v_3$ and $v_5$ in $X$ are mapped to the monomials
$a_3y_3=\frac{1}{{M_3}^6}y_3$ and $a_5y_5=\frac{1}{{M_5}^{12}}y_5$ in the
hypersurface equation (\ref{mirror}) defining $\tilde{X}$. Therefore,
taking the limit $M_3,M_5 \rightarrow \infty$ can be interpreted as
extracting the gauge theory data encoded in the $E_6$ singularity, and
ignoring the region far from the singularity.

In toric geometry language, removing the two vertices $v_3$
and $v_5$ amounts to decompactifying the $K3$ fiber of $X$. Indeed, as
we depict in Figure 1, the ${\bf WP^2_{1,2,3}}$ bundle over ${\bf P^1}$
which is described by $v_1,\cdots,v_5$, is decompactified into ${\bf
C^3}$ which is described by $v_1,v_2$, and $v_4$.
\footnote{
We examined here only the toric data for the ${\bf WP^2_{1,2,3}}$ bundle
over ${\bf P^1}$, not for the entire toric variety, the ${\bf
WP^2_{1,2,3}}$ bundle over ${\bf F_n}$. However, it is straightforward
to check that the entire toric variety is actually decompactified into a
${\bf C^3}$ bundle over ${\bf P^1}$, if the asymptotic freedom condition
$n < -2$ holds.
}
\begin{figure}
   \centerline{\psfig{figure=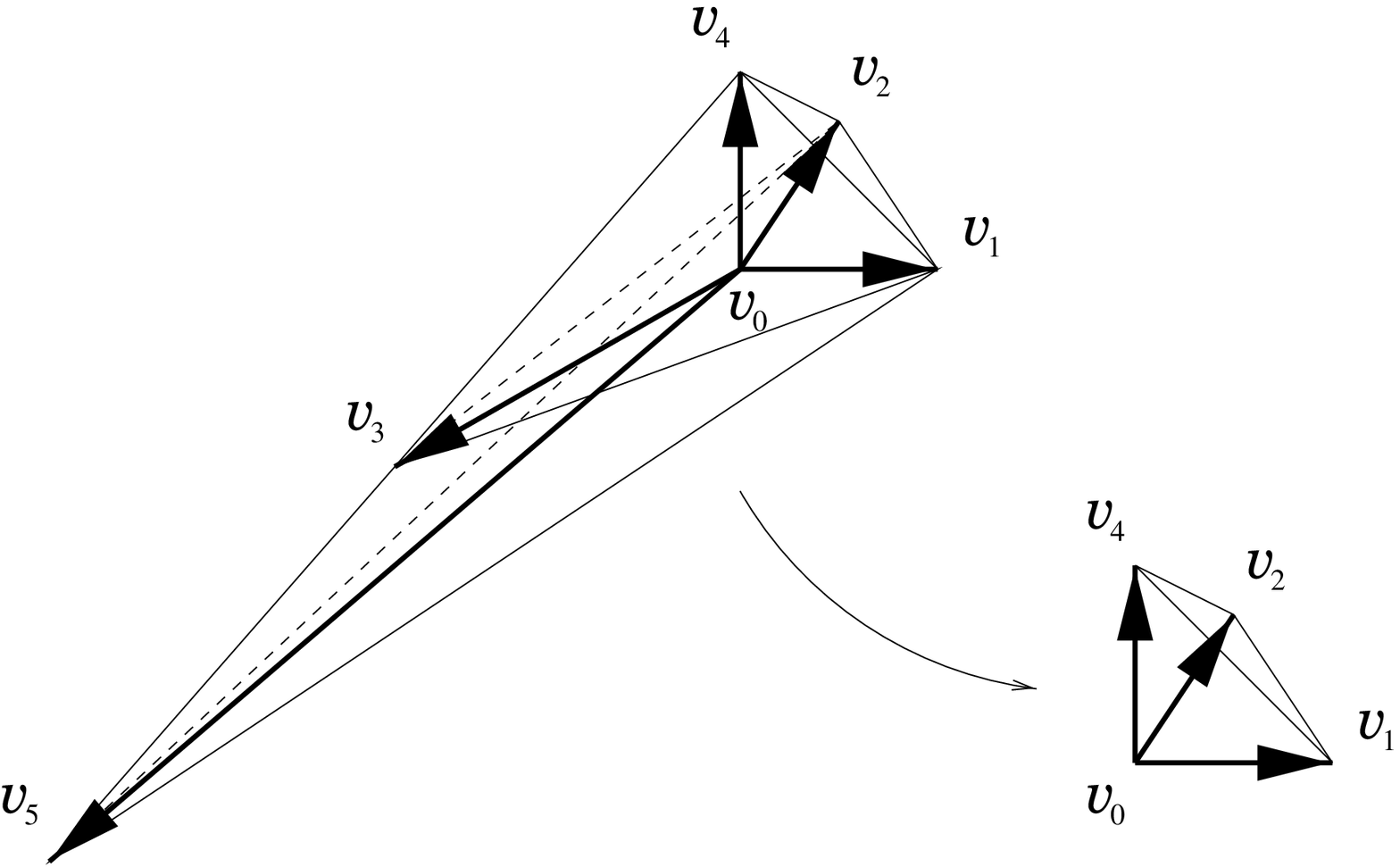,width=11cm}}
   \caption{Decompactification of ${\bf WP^2_{1,2,3}}$ bundle over ${\bf P^1}$.}
\end{figure}
Since the $K3$ fiber is
holomorphically embedded in the ${\bf WP^2_{1,2,3}}$ bundle, the $K3$ is
also decompactified in this process. The divisors $\{z=0\}$ and
$\{t=0\}$ are sent far away from the $E_6$ singularity. The precise
meaning of the terminology ``decompactification'' used here is that the
volumes of the fiber and the base of the elliptic $K3$ surface become
relatively large compared to those of the exceptional ${\bf
P^1}$'s. Under this condition, while the masses of the ``W bosons''
corresponding to D2-branes wrapped on the exceptional ${\bf P^1}$'s are
kept finite, we can decouple the undesirable fields coming from
D2-branes wrapped on the fiber and the base of the elliptic $K3$ surface.

However, infinitely enlarging the $K3$ is not enough for gravity and
stringy effects to be properly decoupled. In order to completely switch off
gravity and stringy effects, we must take the point particle limit $l_s
\rightarrow 0$, where $l_s$ is the string length. Because gauge fields
live on six dimensional space (four dimensional flat space times the
${\bf P^1}$ base of ${\bf F_n}$), the four dimensional gauge coupling at
the string scale, $g_4$, behaves as
\begin{equation}
  \label{coupling}
  {\ds \frac{1}{{g_4}^2}} \sim \frac{V_b}{{l_s}^2},
\end{equation}
where $V_b$ is the volume of the ${\bf P^1}$ base of ${\bf F_n}$.
\footnote{
More precisely, the volume of the $E_6$ gauge symmetry locus which is
represented by $l^{(3)}$ in (\ref{charge-vectors}).
}
Therefore, we must simultaneously take
$V_b/{l_s}^2 \rightarrow \infty$, because the limit $l_s \rightarrow 0$ has
to be taken so that the strong coupling scale $\Lambda_{E_6} \sim
l_s^{-1}\exp(-1/{g_4}^2)$ remains finite. In general, the complexified
K\"{a}hler structure associated to the curve class $C^a$ in $X$ is
related to the complex parameters present in the type IIB mirror
manifold (\ref{mirror}) as
\begin{equation}
  \label{kahler-complex}
    B^a + iV^a \sim {\ds \frac{{l_s}^2}{2\pi i}}\ln 
                    \left(\prod_{i=0}^{13}a_i^{l_i^{(a)}}\right)~~~~
                    {\rm for}~~~V^a \gg {l_s}^2,
\end{equation}
where $B^a$ denotes NS-NS 2-form background and $V^a$ the volume of the
curve $C^a$. We can therefore determine the behavior of
$V_b$, when it is sufficiently large, as
\begin{equation}
  \label{kahler-complex2}
  \begin{array}{rcl}
    V_b  &=& V^3 \\
        &\sim& {\rm Im}{\ds \frac{{l_s}^2}{2\pi i}}
               \ln \left({\ds \frac{\Lambda_{E_6}}{M_0}}\right)^{24-6N_f},
  \end{array}
\end{equation}
where we have used (\ref{charge-vectors}) and the
correspondence between 2-cycles and the vectors (\ref{charge-vectors}) given 
in footnote \ref{identification} (also recall that
$a_1=a_4=a_6=1$ and $a_2=2$). Comparing
(\ref{coupling}) with (\ref{kahler-complex2}), $M_0$ can be identified
with the string scale $l_s^{-1}$. We thus conclude from
(\ref{kahler-complex2}) that the limit $M_0 \rightarrow \infty$ is
nothing but the point particle limit $l_s \rightarrow 0$.

%%%%%%%%%%%%%%%%%%%%%%%%%%%%%%%%%%%%%%%%%%%%%%%%%%%%%%%%%%%%%%%%%%%
%%%%%%%%%%%%%%%%%%%%%%%%%%%%%%%%%%%%%%%%%%%%%%%%%%%%%%%%%%%%%%%%%%%
\section{$N=2$ $E_7$ gauge theories with fundamentals}
%%%%%%%%%%%%%%%%%%%%%%%%%%%%%%%%%%%%%%%%%%%%%%%%%%%%%%%%%%%%%%%%%%%
%%%%%%%%%%%%%%%%%%%%%%%%%%%%%%%%%%%%%%%%%%%%%%%%%%%%%%%%%%%%%%%%%%%

In this section we consider four dimensional $N=2$ supersymmetric 
$E_7$ gauge theories with massive fundamental hypermultiplets.
Since the fundamental representation ($\bf 56$) of $E_7$ is pseudo-real,
we can consider the case that these matter belong to half
hypermultiplets. 
As in the previous section, we can obtain 
the Seiberg-Witten geometries for these theories 
using the toric data for $E_7$ \cite{BIKMSV}.
The point particle limit for this case is rather trivial 
than the $E_6$ case.
The result with $2 N_f$ half hypermultiplets ($N_f \in
\frac{1}{2}{\bf Z}$) are
\beq
  \label{swe7}
    \zeta + \frac{1}{\zeta}
(\Lambda_{E_7})^{36-12 N_f}( X_{E_7}^{\bf 56} )^{N_f}
    + W_{E_7}(x_1,x_2,x_3;w) = 0,
\eeq
where $X_{E_7}^{\bf 56}={x_2}^2$ and
\beq
  \label{e7_deformation}
  \begin{array}{rcl}
    W_{E_7}(x_1,x_2,x_3;w) &=& {x_3}^2 +{x_2}^3 +{x_2} {x_1}^3
                       + w_2 {x_2}^2 {x_1} + w_6 {x_2}^2 \\
                   & & + w_8 {x_2} {x_1} + w_{10} {x_1}^2
                       + w_{12} {x_2} + w_{14} {x_1}+ w_{18}.
  \end{array}
\eeq
The geometry for $N_f=1/2$ case has been obtained in \cite{Brodie}.
The deformation parameters $w_i$ are written in terms of
the Casimirs invariants constructed from an $N=1$ adjoint chiral
multiplet $\Phi$,
and the explicit relation between them
is given in \cite{NoTeYa}.
Note that in \cite{NoTeYa} the polynomial of the Weierstrass form
\beq
\tilde{W}_{E_7}(x,y,z;\tilde{w}) = {y}^2 +{x}^3 
                       + (z^3+ \tilde{w}_8 z+ \tilde{w}_{12} ) x 
                       + \tilde{w}_2 z^4 + \tilde{w}_6 z^3 
                       +\tilde{w}_{10} z^2 + \tilde{w}_{14} z+ \tilde{w}_{18}
\eeq
is used to define the deformation parameters.
This is equivalent to $W_{E_7}(x_1,x_2,x_3;w)$ by the coordinate change
\beq
\begin{array}{rcl}
y &=& {x_3}, \\
x &=& {x_2}+{\ds \frac{1}{3}} {x_1} w_2 + {\ds \frac{1}{3}} w_6, \\
z &=& {x_1}-{\ds \frac{1}{9}} w_2^2,
\end{array}
\eeq
and the redefinition of the Casimirs
\beq
\begin{array}{rcl}
\tilde{w}_2 &=& -{\ds \frac{1}{3}} w_2, \\
\tilde{w}_6 &=& -{\ds \frac{1}{3}}\,{  
w_6}-{\ds \frac {2}{27}}\,{{  w_2}}^{3}, \\
\tilde{w}_8 &=& {  w_8
}-{\ds \frac{1}{27}}\,{{  w_2}}^{4}-{\ds \frac{2}{3}}\,{  w_2}\,{  w_6}, \\
\tilde{w}_{10} &=& {  w_{10}}+{\ds \frac{1}{9}}
\,{{  w_2}}^{2}{  w_6}-{\ds \frac{1}{3}}\,{  w_2}\,{  w_8}, \\
\tilde{w}_{12} &=& {  w_{12}}-{\ds \frac {2}{
729}}\,{{  w_2}}^{6}+{\ds \frac{1}{9}}\,{{  w_2}}^{2}{  w_8}-{\ds \frac {2}{27}}\,
{{  w_2}}^{3}{  w_6}-{\ds \frac{1}{3}}\,{{  w_6}}^{2}, \\
\tilde{w}_{14} &=& {  w_{14}}+{\ds \frac{1}{27}}\,{{  w_2}}^{4}
 {  w_6}+{\ds \frac {2}{2187}}\,
{{  w_2}}^{7}+{\ds \frac{2}{9}}\,{{  w_2}}^{2}{  w_{10}} \\
&& -{\ds \frac {2}{27}}\,{{  
w_2}}^{3}{  w_8}+{\ds \frac{2}{9}}\,{  w_2}\,{{  w_6}}^{2}-{\ds \frac{1}{3}}\,{  w_2}\,{
  w_{12}}-{\ds \frac{1}{3}}\,{  w_6}\,{  w_8}, \\
\tilde{w}_{18} &=&  {  w_{18}}+{\ds \frac {2}{27}}\,{{  w_6}}^{3}-{\ds \frac{1}{3}}\,{
  w_6}\,{  w_{12}}+{\ds \frac {5}{2187}}\,{{  w_2}}^{6}{  w_6}+{\ds
\frac {1}{19683}}\,{{  w_2}}^{9}+{\ds \frac{1}{9}}\,{{  w_2}}^{2}{  w_{14}} \\
&& +{\ds \frac {1}{81}}\,{{  w_2}}^{4}{  w_{10}}-{\ds \frac {1}{243}}\,{{  w_2}
}^{5}{  w_8}+{\ds \frac {2}{81}}\,{{  w_2}}^{3}{{  w_6}}^{2}-{\ds \frac{1}{27}}\,{
{  w_2}}^{3}{  w_{12}}-{\ds \frac{1}{27}}\,{  w_6}\,{{  w_2}}^{2}{  w_8}.
\end{array}
\eeq

If we consider the massive hypermultiplet,
only the polynomial of degree twelve $X_{E_7}^{\bf 56}$ 
should be modified in (\ref{swe7}).
Below we will determine $X_{E_7}^{\bf 56}(x,y,z;w,m)$ for the massive case.
By giving appropriate VEV's to the moduli, the geometry (\ref{swe7}) 
should reduce to the one describing $E_6$ or $SO(12)$ gauge theory 
as in \cite{TeYa2}.
\footnote{$SU(7)$ gauge theory 
can also be obtained in this way. 
But we will not consider this case because actual computation is very
difficult.}

First, we consider the reduction to the $E_6$ gauge theory
by removing the simple root $\alpha_6$.
For the notation for roots and weights we follow \cite{Slansky} 
and \cite{TeYa2}.
According to \cite{TeYa2},
by tuning the Higgs vector as
$ a^i =(M+\delta a^1, 2 M+\delta a^2, 3 M+\delta a^3, 
\frac{5}{2} M+\delta a^4, 2M+\delta a^5, \frac{3}{2} M,
\frac{3}{2} M+\delta a^6)$
and taking the limit $M \rightarrow \infty$,
we should obtain the $E_6$ singularity.
Indeed by explicit calculations,
we find that
\beq
\tilde{W}_{E_7}(x,y,z;\tilde{w}) = 
\left( 2 M \right)^6 W_{E_6}(x',y',z' ;w'(\delta a_i))+{\cal O}(M^5),
\eeq
where
\beq
\begin{array}{rcl}
x &=& 4 M^2 x', \\
y &=& 8 M^3 y', \\
z &=& -4 M z'-{\ds \frac{1}{2}} M^2(\tilde{w}_2-{\ds \frac{1}{4}} M^2),
\end{array}
\eeq
and $W_{E_6}$ is given in (\ref{e6_deformation}).
The fundamental representation ${\bf 56}$ of $E_7$ is decomposed into
the representation of $E_6 \times U(1)$ as
\beq
{\bf 56} ={\bf 27}_{\frac{1}{2}} \oplus 
\overline{{\bf 27}}_{-\frac{1}{2}} \oplus {\bf 1}_{\frac{3}{2}} \oplus {\bf 1}_{-\frac{3}{2}},
\eeq
where the subscript denotes the $U(1)$ charge $\alpha_6 \cdot \lambda_i$.
Thus if we take $m=-\frac{1}{2} M + m'$,
we should obtain the Seiberg-Witten geometry for the $E_6$ theory with
fundamental matter in the limit $M \rightarrow \infty$.
This means that
\beq
\label{cond1}
X_{E_7}^{\bf 56}(x,y,z;w,m) = C M^6 X_{E_6}^{\bf 27}(x',y',z';w',m')+
{\cal O}(M^5),
\eeq
where $X_{E_6}^{\bf 27}$ is proposed in \cite{TY} as
\beq
\label{e6fund}
X_{E_6}^{\bf 27}(x,y,z;w,m)=m^6+2 w_2 m^4-8 m^3 z+({w_2}^2-12 x)m^2
+ 4 m w_5  -4 w_2 x - 8(z^2-i y+w_6/2),
\eeq
and $C$ is a constant.

Next we consider the gauge symmetry breaking which yields the $SO(12)$ gauge
theory with spinors, by giving the VEV
$a^i =(2 M, 3 M+\delta a^5, 4 M+\delta a^4, 
3 M+\delta a^3, 2M+\delta a^2, M+\delta a^1,
2 M+\delta a^6)$ to $\Phi$.
We substitute this into $w_i(a_i)$ and look for the coordinates 
which eliminate the terms of the order of $M^l$ $(9 \leq l \leq 18)$ 
in $W_{E_7}$.
We can find such coordinates as
\beq
\label{coord1}
\begin{array}{rcl}
x &=& {\ds \frac{1}{135}} M^6+M^4 \left( {\ds \frac{1}{3}} z'+
{\ds \frac{1}{30}} \tilde{w}_2 \right) + M^2 \left( {\ds \frac{1}{2}} i x'+
{\ds \frac{1}{10}} \tilde{w}_2^{~2} \right)-{\ds \frac{1}{10}} \tilde{w}_6, \\
y &=& {\ds \frac{1}{2}} i M^4 y', \\
z &=& -{\ds \frac{1}{3}} M^4
      +M^2 \left( \tilde{w}_2 -{\ds \frac{1}{2}} z'\right)
      +{\ds \frac{3}{2}} z'^2-{\ds \frac{3}{2}} \tilde{w}_2 z'
      -\tilde{w}_2^{~2},
\end{array}
\eeq
in terms of which the polynomial $W_{E_7}$ describing the $E_7$
singularity is represented as 
\beq
W_{E_7} (x,y,z;\tilde{w}) = -4 M^8 W_{D_6} (x',y',z';v) +{\cal O}(M^7),
\eeq
where 
\beq
 W_{D_6} (x,y,z;w)=y^2+ z x^2+z^5 +v_2 z^4+ v_4 z^3+v_6 z^2++v_8
z+v_{10}+2 i x {\rm Pf},
\eeq
and $v_i$ and ${\rm Pf}$ are Casimirs of $SO(12)$ built out of $\delta a_i$.
The explicit forms of them can be read off from
\beq
W_{D_6} (x,y,z;w)=y^2+z x^2+\frac{1}{z} \left( \prod_{i=1}^{6} (z-{b_i}^2)
-\prod_{i=1}^{6} {b_i}^2 \right) +2 i x \prod_{i=1}^{6} b_i,
\eeq
where $b_1=\delta a_1,b_2=\delta a_2-\delta a_1, 
b_3=\delta a_3-\delta a_2,b_4=\delta a_4-\delta a_3,
b_5=\delta a_5 +\delta a_6-\delta a_4$ and $b_6=\delta a_6-\delta a_5$.
The fundamental representation ${\bf 56}$ of $E_7$ is decomposed into
the representation of $SO(12) \times U(1)$ as
\beq
{\bf 56} ={\bf 32}_{0} \oplus 
{\bf 10}_{1} \oplus \overline{{\bf 10}}_{-1},
\eeq
where the subscript denotes the $U(1)$ charge $\alpha_1 \cdot \lambda_i$.
The indices of spinor representation ${\bf 32}$ and fundamental
representation ${\bf 10}$ are eight and two, respectively.
Thus the terms proportional to $M^l(5 \leq l \leq 12)$ in $X_{E_7}^{\bf 56}$
must be absent after taking the coordinates $(x'_1,x'_2,x'_3)$ defined
in (\ref{coord1}).
Note that there is no need to shift the mass to make the spinor matter 
survive. 
Although this and the condition (\ref{cond1}) impose
very strong restrictions on the polynomial $X_{E_7}^{\bf 56}(x,y,z;w,m)$,
we can find a unique solution. In the coordinates $(x_1,x_2,x_3)$ it is
\beq
\label{pol7}
\begin{array}{rcl}
X_{E_7}^{\bf 56} (x_1,x_2,x_3;w,m) &=&
{m}^{12}+2\,{  w_2}\,{m}^{10}+\left (6\,{  x_1}+{{  w_2}}^{2}
\right ){m}^{8} \\
&& +\left (2\,{  x_1}\,{  w_2}-10\,{  x_2}-4\,{  w_6
}\right ){m}^{6}+\left (-3{{  x_1}}^{2}  -6{  w_2}{  x_2}-4\,{
  w_8}\right ){m}^{4} \\
&& +8\,i {  x_3}{m}^{3}+\left (-6\,{
  x_2}\,{  x_1}-4 {  w_{10}}\right ){m}^{2}+{ x_2}^{2},
\end{array}
\eeq
and $C=1$.
Note that if we take $m=0$, $X_{E_7}^{\bf 56}$ becomes a factorized form
and agrees with the massless case obtained from Calabi-Yau construction.
In the semi-classical limit $\Lambda_{E_7} \rightarrow 0$,
the low energy theory has singularities associated with
massless squarks \cite{TY} when 
\beq
0=\Delta_M (m;w_i) \equiv
{\rm det}_{56 \times 56} (m {\bf 1 }-\Phi^{cl})=
m^{56}+432 w_2 m^{54}+ \cdots .
\eeq
In fact, the hypersurface defined by the intersection of 
$X_{E_7}^{\bf 56}(x_1,x_2,x_3;w,m)=0$ 
and $W_{E_7}(x_1,x_2,x_3;w)=0$ in ${\bf C^3}$ parametrized by
$(x_1,x_2,x_3)$ becomes singular
when $w_i$'s satisfy $\Delta_M =0$.
This means that in the limit $\Lambda_{E_7} \rightarrow 0$ the
Seiberg-Witten geometry becomes singular when $\Delta_M =0$.
This fact can be regarded as a non-trivial check of the validity of
(\ref{pol7}). It is straightforward to generalize (\ref{swe7}) to the
theory with fundamental hypermultiplets with different masses.

We can also find the Seiberg-Witten geometry for $N=2$ $SO(12)$ gauge
theory with spinor matter from (\ref{pol7}) and (\ref{coord1}).
Indeed we obtain
\beq
X_{E_7}^{\bf 56}(x_1,x_2,x_3;w;m) = M^4 X_{D_6}^{\bf
32}(x',y',z';v;m)+{\cal O}(M^3),
\eeq
where 
\beq
\label{d6spinor}
\begin{array}{rcl}
X_{D_6}^{\bf 32}(x,y,z;v;m) &=& {m}^{8}+\left (2\,z+{  v_2}\right ){m}^{6}
+\left (3 i \,x+{\ds \frac{1}{2}}\,{  
v_2}\,z+{\ds \frac{3}{8}}\,{{  v_2}}^{2}
-{\ds \frac{1}{2}} \,{  v_4}\right ){m}^{4}-4 i\,y{m}^{3} \\
&& +\left (-2\,{  {\rm Pf}}+{  v_6}-\! {\ds \frac{1}{4}}\,{  v_4}\,
     {  v_2}+\! {\ds \frac{1}{16}}\,{{  v_2}}^
{3}+2\,{  v_2}\,{z}^{2} \right. \\
&& \left. -\! {\ds \frac{1}{8}}\,z{{  v_2}}^{2}+\! {\ds \frac{3}{2}}\,z{  v_4}+3\,{z}^{3
}+\! 3 i \,zx+\! {\ds \frac{1}{2}}i \,x{  v_2}\right ){m}^{2} \\
&& +{\! {\ds \frac {1}{256}}}\,\left (8 i \,x+
8\,{z}^{2}+4\,{  v_2}\,z+4\,{  v_4}-{{  v_2}}^{2}\right )^{2}.
\end{array}
\eeq
Taking $m=-M+m'$, which corresponds to the reduction 
to the $SO(12)$ gauge theory 
with fundamental matter, we also obtain 
\beq
X_{E_7}^{\bf 56}(x_1,x_2,x_3;w;m) = 4 M^{10} (m'^2-z')+{\cal O}(M^9).
\eeq
Therefore the Seiberg-Witten geometry for the $N=2$ $SO(12)$ gauge
theory with $N_s$ spinors and $N_f$ fundamentals is 
\beq
  \label{swso12}
    \zeta + \frac{1}{\zeta}
(\Lambda_{D_6})^{20-8 N_s-2 N_f} \prod_{i=1}^{N_s} 
 X_{D_6}^{\bf 32}(x,y,z;v;m_i) 
\prod_{j=1}^{N_f} ( {m'}_j^2 - z)
    + W_{D_6}(x,y,z;v) = 0.
\eeq
Note that in the massless case,
the polynomial $X_{D_6}^{\bf 32} $ is factorized and 
agrees with that obtained in \cite{AG}.
It seems that 
the polynomials (\ref{pol7}) and (\ref{d6spinor}) 
cannot be derived from confining phase superpotential technique
because they have the term of odd degrees in $m$.

%%%%%%%%%%%%%%%%%%%%%%%%%%%%%%%%%%%%%%%%%%%%%%%%%%%%%%%%%%%%%%%%%%%
%%%%%%%%%%%%%%%%%%%%%%%%%%%%%%%%%%%%%%%%%%%%%%%%%%%%%%%%%%%%%%%%%%%
\section{Discussion and conclusions}
%%%%%%%%%%%%%%%%%%%%%%%%%%%%%%%%%%%%%%%%%%%%%%%%%%%%%%%%%%%%%%%%%%%
%%%%%%%%%%%%%%%%%%%%%%%%%%%%%%%%%%%%%%%%%%%%%%%%%%%%%%%%%%%%%%%%%%%

In this article, we have studied $N=2$ supersymmetric $E_6$ and $E_7$ gauge
theories with fundamental matter. For the $E_7$ theory, we have taken
various decoupling limits such that the $E_7$ theory flows to the theories
with smaller gauge groups, the exact solutions of which have been
already known. As a result, the $E_7$ geometry proposed in section 3
was shown to reduce exactly to the geometries expected from earlier
works. This observation serves as a non-trivial check of the validity
of the original $E_7$ geometry. Furthermore, by breaking the group $E_7$
appropriately, we have also found the Seiberg-Witten geometry for
$SO(12)$ gauge theories with massive fundamentals and spinors.

We have analyzed the $E_6$ gauge theories, by realizing them as
decoupled theories geometrically contained in type IIA string theory
compactified on singular Calabi-Yau threefolds. Although the
Seiberg-Witten geometry for this theory has been already constructed
from field theoretical technique, it is meaningful to ascertain whether
or not the same geometry can be reproduced from local structure of
Calabi-Yau manifolds. Remarkably, we have found a complete agreement in
the case of massless fundamental matter.

One of the disadvantages in using Calabi-Yau manifolds to determine the
geometry is that the masses for matter multiplets cannot be easily
incorporated. As mentioned in section 2, the matter multiplets are
constrained on the extra singularities on the ${\bf P^1}$ base of
Calabi-Yau manifolds. In order to predict how the hypermultiplet masses
modify the Seiberg-Witten geometry from geometrical point of view, we
presumably need to blow up the Calabi-Yau at the extra singularities. At
first sight, however, it seems extremely difficult to account for the
intricate mass dependence such as (\ref{e6fund}), (\ref{pol7}), and
(\ref{d6spinor}).

Calabi-Yau construction of gauge theories has another disadvantage that
we cannot realize gauge groups with arbitrarily large ranks. To overcome 
this difficulty, we require a more powerful framework, ``geometric
engineering'' \cite{KKV}, which embeds gauge theories in non-compact
Calabi-Yau threefolds instead of compact ones. Geometric engineering is
presumably the most systematic and general approach to the problem of
finding Seiberg-Witten geometries, even though there are several other
methods which make use of branes, such as brane probe and MQCD.
It is probable that understanding full geometrical features of string
theories will enable us to investigate $N=2$ supersymmetric theories with
{\it arbitrary} gauge groups and matter representations.

%%%%%%%%%%%%%%%%%%%%%%%%%%%%%%%%%%%%%%%%%%%%%%%%%%%%%%%%%%%%%%%%%%%
\section*{Acknowledgements}
%%%%%%%%%%%%%%%%%%%%%%%%%%%%%%%%%%%%%%%%%%%%%%%%%%%%%%%%%%%%%%%%%%%

We would like to thank T. Tani and K. Hosomichi for valuable
discussions. The work of J.H. and S.T. was supported by JSPS Research
Fellowships for Young Scientists.

\clearpage

%%%%%%%%%%%%%%%%%%%%%%%%%%%%%%%%%%%%%%%%%%%%%%%%%%%%%%%%%%%%%%%%%%%
%%%%% ** Reference ** %%%%%%%%%%%%%%%%%%%%%%%%%%%%%%%%%%%%%%%%%%%%%
%%%%%%%%%%%%%%%%%%%%%%%%%%%%%%%%%%%%%%%%%%%%%%%%%%%%%%%%%%%%%%%%%%%


\begin{thebibliography}{99}

%%%%%%%%%%%%%%%%%%%%%%%%%%%%%%%%%%%%%%%%
\bibitem{SW}
  N. Seiberg and E. Witten,
  Nucl. Phys. {\bf B426} (1994) 19, Erratum-ibid. {\bf B430} (1994) 485,
  hep-th/9407087;
  Nucl. Phys. {\bf B431} (1994) 484, hep-th/9408099.
%%%%%%%%%%%%%%%%%%%%%%%%%%%%%%%%%%%%%%%%
\bibitem{KMP}
  S. Katz, D.R. Morrison and M.R. Plesser,
  Nucl. Phys. {\bf B477} (1996) 105, hep-th/9601108.
%%%%%%%%%%%%%%%%%%%%%%%%%%%%%%%%%%%%%%%%
\bibitem{KKV}
  S. Katz, A. Klemm and C. Vafa,
  Nucl. Phys. {\bf B497} (1997) 173, hep-th/9609239.
%%%%%%%%%%%%%%%%%%%%%%%%%%%%%%%%%%%%%%%%
\bibitem{KMV}
  S. Katz, P. Mayr and C. Vafa,
  Adv. Theor. Math. Phys. {\bf 1} (1998) 53, hep-th/9706110.
%%%%%%%%%%%%%%%%%%%%%%%%%%%%%%%%%%%%%%%%
\bibitem{Mayr}
  For an elementary review, see
  P. Mayr,
  Fortsch. Phys. {\bf 47} (1999) 39, hep-th/9807096.     
%%%%%%%%%%%%%%%%%%%%%%%%%%%%%%%%%%%%%%%%
\bibitem{TY}
  S. Terashima and S.-K. Yang,
  Phys. Lett. {\bf B430} (1998) 102, hep-th/9803014.
%%%%%%%%%%%%%%%%%%%%%%%%%%%%%%%%%%%%%%%%
\bibitem{Brodie}
  J.H. Brodie,
  Nucl. Phys. {\bf B506} (1997) 183, hep-th/9705068.     
%%%%%%%%%%%%%%%%%%%%%%%%%%%%%%%%%%%%%%%%
\bibitem{KLMVW}
  A. Klemm, W. Lerche, P. Mayr, C. Vafa and N. Warner,
  Nucl. Phys. {\bf B477} (1996) 746, hep-th/9604034.     
%%%%%%%%%%%%%%%%%%%%%%%%%%%%%%%%%%%%%%%%
\bibitem{LW}
  W. Lerche and N.P. Warner,
  Phys. Lett. {\bf B423} (1998) 79, hep-th/9608183.     
%%%%%%%%%%%%%%%%%%%%%%%%%%%%%%%%%%%%%%%%
\bibitem{AG}
  M. Aganagic and M. Gremm,
  Nucl. Phys. {\bf B524} (1998) 207, hep-th/9712011.
%%%%%%%%%%%%%%%%%%%%%%%%%%%%%%%%%%%%%%%%
\bibitem{HKTY}
  S. Hosono, A. Klemm, S. Theisen and S.-T. Yau,
  Commun. Math. Phys. {\bf 167} (1995) 301, hep-th/9308122;
  Nucl. Phys. {\bf B433} (1995) 501, hep-th/9406055.
%%%%%%%%%%%%%%%%%%%%%%%%%%%%%%%%%%%%%%%%
\bibitem{Batyrev}
  V.V. Batyrev,
  J. Alg. Geom. {\bf 3} (1994) 493, alg-geom/9310003.
%%%%%%%%%%%%%%%%%%%%%%%%%%%%%%%%%%%%%%%%
\bibitem{AGR}
  P.S. Aspinwall, B.R. Greene and D.R. Morrison,
  Internat. Math. Res. Notices (1993) 319, alg-geom/9309007.
%%%%%%%%%%%%%%%%%%%%%%%%%%%%%%%%%%%%%%%%
\bibitem{Greene}
  B.R. Greene,
  ``String Theory on Calabi-Yau Manifolds'',
  in {\it Fields, Strings and Duality, TASI 1996},
  C. Esthimiou and B.R. Greene (eds),
  World Scientific (1997) New Jersey, 
  hep-th/9702155.
%%%%%%%%%%%%%%%%%%%%%%%%%%%%%%%%%%%%%%%%
\bibitem{KachVa}
  S. Kachru and C. Vafa,
  Nucl.Phys. {\bf B450} (1995) 69, hep-th/9505105.     
%%%%%%%%%%%%%%%%%%%%%%%%%%%%%%%%%%%%%%%%
\bibitem{Aspinwall}
  P.S. Aspinwall,
  ``K3 Surfaces and String Duality'',
  in {\it Fields, Strings and Duality, TASI 1996},
  C. Esthimiou and B.R. Greene (eds),
  World Scientific (1997) New Jersey,
  hep-th/9611137.     
%%%%%%%%%%%%%%%%%%%%%%%%%%%%%%%%%%%%%%%%
\bibitem{BIKMSV}
  M. Bershadsky, K. Intriligator, S. Kachru, D.R. Morrison, V. Sadov and 
 C. Vafa,
  Nucl. Phys. {\bf B481} (1996) 215, hep-th/9605200.     
%%%%%%%%%%%%%%%%%%%%%%%%%%%%%%%%%%%%%%%%
\bibitem{KV}
  S. Katz and C. Vafa,
  Nucl. Phys. {\bf B497} (1997) 146, hep-th/9606086.
%%%%%%%%%%%%%%%%%%%%%%%%%%%%%%%%%%%%%%%%
\bibitem{Tani}
  T. Tani,
  to appear.
%%%%%%%%%%%%%%%%%%%%%%%%%%%%%%%%%%%%%%%%
\bibitem{CF}
  P. Candelas and A. Font,
  Nucl. Phys. {\bf B511} (1998) 295, hep-th/9603170.
%%%%%%%%%%%%%%%%%%%%%%%%%%%%%%%%%%%%%%%%
\bibitem{ElFoGiInRa} S. Elitzur, A. Forge, A. Giveon, K. Intriligator and 
E. Rabinovici, Phys. Lett. {\bf B379} (1996) 121, hep-th/9603051.
%%%%%%%%%%%%%%%%%%%%%%%%%%%%%%%%%%%%%%%%
\bibitem{KiTeYa}
T. Kitao, S. Terashima and S.-K. Yang,
Phys. Lett. {\bf B399} (1997) 75, hep-th/9701009. 
%%%%%%%%%%%%%%%%%%%%%%%%%%%%%%%%%%%%%%%%
\bibitem{TeYa1} 
S. Terashima and S.-K. Yang, Nucl. Phys. {\bf B519} (1998) 453,
hep-th/9706076.
%%%%%%%%%%%%%%%%%%%%%%%%%%%%%%%%%%%%%%%%
\bibitem{NoTeYa}
  M. Noguchi, S. Terashima and S.-K. Yang,
  Nucl. Phys. {\bf B556} (1999) 115, hep-th/9903215.
%%%%%%%%%%%%%%%%%%%%%%%%%%%%%%%%%%%%%%%%
\bibitem{TeYa2}
  S. Terashima and S.-K. Yang,
  Nucl. Phys. {\bf B537} (1999) 344, hep-th/9808022.
%%%%%%%%%%%%%%%%%%%%%%%%%%%%%%%%%%%%%%%%
\bibitem{Slansky}
  R. Slansky,
  ``Group Theory for Unified Model Building'',
  Phys. Rep. {\bf 79} (1981) 1.
%%%%%%%%%%%%%%%%%%%%%%%%%%%%%%%%%%%%%%%%

\end{thebibliography}
\end{document}